# Contravening Esotery: Cryptanalysis of Knapsack Cipher using Genetic Algorithms

Harmeet Singh
Software Engineer
CEB Global, Gurgaon
Haryana, India

## ABSTRACT
Cryptanalysis of knapsack cipher is a fascinating problem which has eluded the computing fraternity for decades. However, in most of the cases either the time complexity of the proposed algorithm is colossal or an insufficient number of samples have been taken for verification. The present work proposes a Genetic Algorithm based technique for cryptanalysis of knapsack cipher. The experiments conducted prove the validity of the technique. The results prove that the technique is better than the existing techniques. An extensive review has been carried out in order to find the gaps in the existing techniques. The work paves the way of the application of computational intelligence techniques to the discipline of cryptanalysis.

## 1. INTRODUCTION
The word cryptography means secret writing. The discipline includes encryption of data and its decryption. However the secrecy of data did not become a discipline itself until the Second World War. The agencies of the United States of America developed cryptographic systems before the Second World War and there was hardly any open source literature available on the techniques of cryptography. The discipline however got a boost by the accidental declassification of related documents after the world war [1]. Cryptography has seen many local maximas and it is not certain that the present algorithms are good enough to stand the attacks of hackers. It is therefore required to amalgamate the Computational Intelligence techniques with the existing systems in order to make data more secure and still achieve the target of minimum complexity. It is also required to develop cryptanalysis techniques for the existing systems in order to analyse the techniques which can break a cipher easily. The technique presented in the present paper henceforth referred to as "the work" accomplishes the task of cryptanalysis of knapsack ciphers. The technique is based on Genetic Algorithms (GAs).

Knapsack algorithm was one of the first algorithms of public key encryption. The initial version of the algorithm could only be used for encryption [1]. However, later on it was modified in order to perform decryption as well.

The Knapsack cipher is a depiction of an NP complete problem. An instance of Knapsack Problem having lesser number of elements can be solved easily. However, solving a bigger set will be of exponential complexity. GAs can, therefore, be used in order to find out the solution of a knapsack problem. Since GAs are known for achieving optimization. The encryption part of the algorithm is purely mathematical in nature. However for the decryption, heuristic search algorithms are required. GAs have therefore been used in order to find the original message from the encrypted key.

There are two versions of the problem: the harder and the easier. The harder version of knapsack problem has been derived from the easier version. The harder version is used to create a public key. Though, it has been shown by Schneier [1] that knapsack system can be broken, however, it was only in certain cases that the analysis proved correct. The communication of the key was dealt with by Shamir [2]. It may be noted that the effect of variations in the operators of GAs have not been analyzed in detail as yet.

The present work shows the effect of variation of cross over rate, mutation rate and the type of cross over on the strength of the key. The work has the following contributions

1). To present a novel technique of the cryptanalysis of the knapsack cipher using computational intelligence methods.

2). To study the effect of variation of parameter on the ability of the algorithm to find the solution.

The rest of the paper has been organized as follows. Section 2 presents the literature review, Section 3 explains the concept of GAs, Section 4 presents the proposed algorithm, Section 5 discusses its results & the last section discusses the conclusions. The proposed algorithm paves the way of Computational Intelligence techniques in the field of cryptanalysis.

## 2. LITERATURE REVIEW
The section deals with the need of the review, the research questions and elaborates the sources of information along with the search criteria, the selection of studies, the methodology of data extraction and finally the threats of validity.

The papers were filtered on the basis of titles. This was followed by segregation on the basis of keywords. Then the abstract of the remaining papers were studied. The final set of papers was selected by reading the complete text. The final set has 18 relevant papers. The papers on knapsack cipher cryptanalysis were further segregated on the basis of type of technique. Finally, 18 papers were selected.

**Table 1. Review Of Cryptanalysis Knapsack Cipher**

| S.No. | Name of the author | Reference Number | Summary |
|---|---|---|---|
| 1 | Sinha, S.N., Palit, S. et. al. | 5 | The use of Differential Evolution which is a population based search strategy using crossover, mutation and selection for the cryptanalytic attack on Merkle-Hellman Knapsack |
| | | | cipher is presented in the paper. |
| 2 | Raghuvamshi, A. Rao, P.V. | 6 | The simple mathematics to break knapsack cipher for cryptanalysis on knapsack cipher is used in this paper. |
| 3 | Palit, S., Sinha, S.N., et. al. | 7 | The decryption if cipher text to plaintext is presented in this paper for cryptanalysis of knapsack cipher using Firefly Algorithm(FA). |
| 4 | AbdulHalim, M.F., Hameed, S.M., et. al. | 8 | For the cryptanalysis of Knapsack Cipher the binary particle swarm optimization for deducing the plaintext from cipher test is presented in this paper. |
| 5 | N. Nalini and G. Raghavendra Rao | 9 | The successful attacks on some block ciphers which are delegate in computer security is presented in this paper. |
| 6 | N. Nalini and G. Raghavendra Rao | 10 | Simplified Data Encryption Standard (SDES) and Modified version of Data Encryption Standard (DES) and the attacks are presented in the paper. |
| 7 | Jun Song China, Huanguo Zhang, et. al. | 11 | Genetic Algorithm on Two-round DES system is presented in this paper. |
| 8 | Jun Song, Huanguo Zhang, Qingshu, et. al. | 12 | The various optimum keys on fitness functions and plaintext attack is presented in this paper. |
| 9 | Bart Preneel | 13 | In this paper cryptographic techniques for protecting the information such as hash functions and digital signature schemes are presented. |
| 10 | William Millan, Andrew Clark, Ed Dawson | 14 | The paper presents the generation of highly nonlinear balanced Boolean functions by genetic algorithms using the hill climbing and other techniques. |
| 11 | Bart Preneel | 15 | The cryptographic hash functions are used to present some problems and most important constructions are described in this paper. |
| 12 | Bart Preneel | 16 | In this paper the brief outline of the hash functions are explained in the starting which is clarified afterwords in the paper. |
| 13 | W. Millan, L. Burnett, G. Carter, et. al. | 17 | The selection of suitable GA parameters, and in particular an effective technique for breeding s-boxes is discussed. |
| 14 | Bart Preneel | 18 | In this paper the brief summary of hash functions 30 years after their introduction and SHA-3 competition progressed are discussed. |
| 15 | Kunikatsu Kobayashi, Masaki Kimura | 19 | The LLL algorithm for analyzing and encryption of keys with the choice in a typical cryptanalysis algorithm is explained and proposed in this paper. |
| 16 | Haitham Rashwan1, Ernst M. Gabidulin 2, et. al. | 20 | In this paper, the security of previously present systems against attacks is increased and furthermore the public key size is reduced to 4Kbits from 10 Kbits is applied. |
| 17 | Rongxing Lu1, Xiaodong Lin2, et. al. | 21 | An efficient IND-CCA2-secure public key encryption scheme is explained in this paper which is based on coding theory. |
| 18 | Bhasin H. et. al. | 22 | Dynamic TSP to compare greedy approach, genetic algorithms and DGAs are presented. |

## 3. GENETIC ALGORITHMS

Genetic algorithms are search techniques based on the mechanics of natural selection and natural genetics [23]. GAs was developed by J. Holland. While developing these algorithms he had many goals in mind, the most important of which was retaining the flair of a natural system. The popularity of these algorithms can be gauged from the fact with the number of papers and dissertations which used GAs and natural search processes have shown a sharp increase after 90's. Their applicability ranged from science to business.

It may be noted that the calculus based search methods solve non linear equations by generally equating gradient to zero. There is another category of calculus based search methods which is based on the notion of steepest hill climbing. However, none of the methods stated so far have the capability of going beyond local maxima. This problem can be tackled by random search which may not be the best computational procedure but overcomes the problems faced in calculus based methods. Random algorithms have become increasingly popular but in the long run there is no guarantee that they perform better than Brute Force algorithms. Cryptography using Genetic Algorithm has already been done[24]. The main steps of GAs are Population Generation, Crossover, Mutation, Fitness evacuation, Selection, Checking condition for termination and Repetition of the above steps.

## 4. PROPOSED WORK

The proposed algorithm finds out the set of constants which would solve the Knapsack Cipher problem.. The process is based on the hybrid genetic algorithms. It may be stated here that the fitness function has been designed keeping in consideration the results obtained in the earlier implementations. Moreover, a sound problem reduction approach has been followed in order to apply GAs to the problem. The steps of the proposed algorithm are as follows.

1. Input Gathering: The input of the algorithm is the elements of the required set and the required sum. The software, designed in Java, asks the user to enter the input. In the experiment the input sets have been deliberately chosen to encompass most of the possible cases.

2. Initial population Generation: The initial population is generated with the help of pseudo random number generator. The number of chromosomes has to be large enough to make crossover and mutation meaningful. For instance if the number of chromosomes is 5 and the crossover rate is two, then the crossover operator will have no effect what so ever. Here, it may be stated that the number of cells in a chromosome is equal to number of elements of the set.

3. Fitness evaluation :

   a) The above step is followed by the multiplication of the elements of the binary chromosome with the elements of the entered set and the calculation of the sum of the values obtained.

   b) This is followed by calculating the difference of the sum calculated in above step and the sum entered as the input in the first step.

   c) The fitness of each chromosome is then calculated using the fitness formula i.e. $fit[i] = (1 / convert[i]) * 100$.

   d) If the calculated difference is equal to zero then the fitness value is taken to be 101 else the fitness is calculated using the above mentioned formula. Here the number 101 signifies maximum fitness.

4. Selection: This is followed by the application of the Roulette Wheel Selection in which the plausibility of the fit chromosome being replicated is high.

5. Crossover and Mutation: The crossover and mutation are then applied on the new population in order to obtain optimized population.

6. Termination: The process stops if either result is achieved or number of generation surpasses the requisite value.

The process has been depicted in Figure 1.

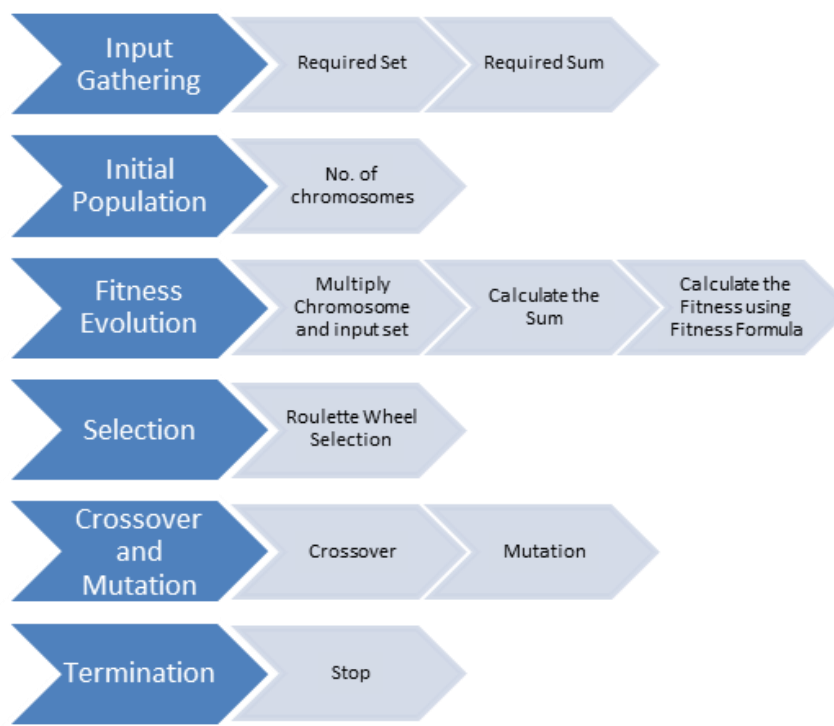

**Fig 1: Graphical depiction of proposed work**

The above process has been implemented and tested. The technique has been applied to other NP Hard Problems like Maximum Clique, Travelling Salesman and Subset Sum problem[25, 26, 27]. The next two sections discuss the results and the analysis of the results obtained.

## 5. RESULTS

The proposed algorithm has been implemented in Java and tested for 5 sets. The crossover was varied from 2 to 5. The mutation rate was varied from 0.5 to 0.8. For each set, the experiment is executed five times for a particular mutation rate and crossover rate. The data of the experiments carried out is as follows.

First Experiment

- Set A1 = {2, 4, 6, 8, 10, 12}
- The value of the sum, m = 20
- Cross over rate: 2

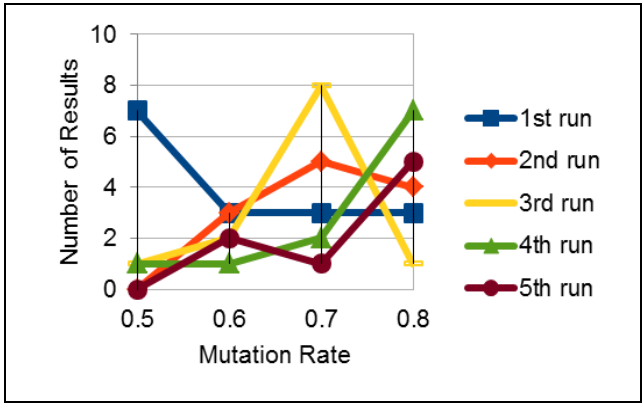

**Fig 2: Results of Experiment 1.**

Second Experiment

- Set A1 = {2, 4, 6, 8, 10, 12}
- The value of the sum, m = 20
- Cross over rate: 3

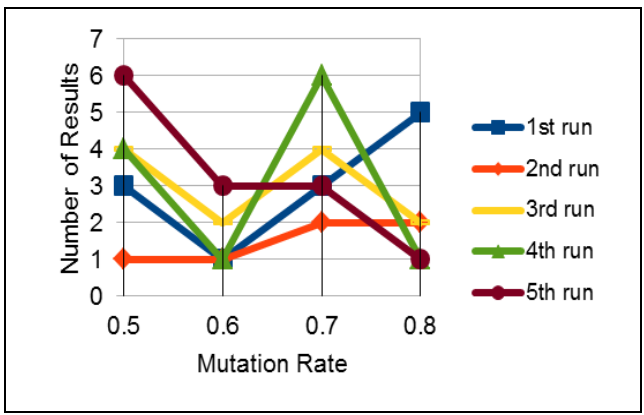

**Fig 3: Results of Experiment 2.**

Third Experiment

- Set A1 = {2, 4, 6, 8, 10, 12}
- The value of the sum, m = 20
- Cross over rate: 4

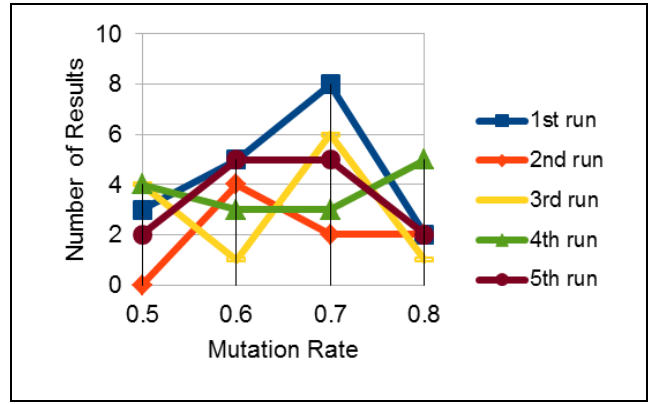

**Fig 4: Results of Experiment 3.**

Fourth Experiment

- Set A1 = {2, 4, 6, 8, 10, 12}
- The value of the sum, m = 20
- Cross over rate: 5

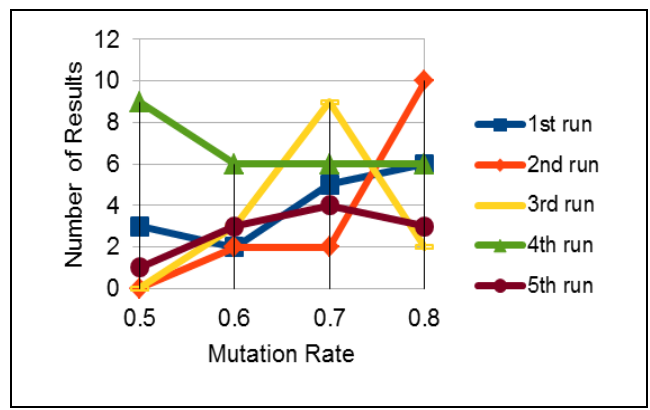

**Fig 5: Results of Experiment 4.**

Fifth Experiment

- Set A1 = {1, 3, 5, 7, 9, 11}
- The value of the sum, m = 20
- Cross over rate: 2

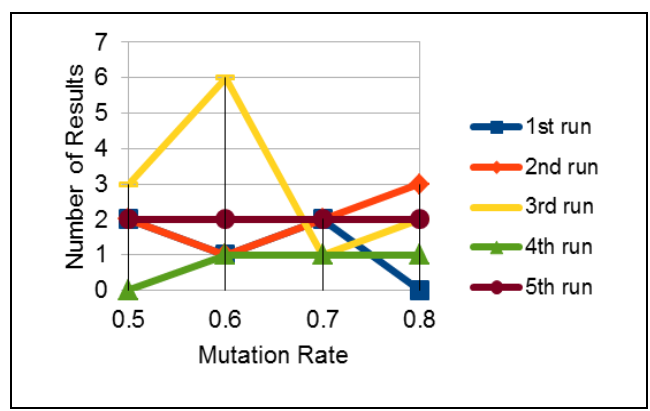

**Fig 6: Results of Experiment 5.**

Sixth Experiment

- Set A1 = {1, 3, 5, 7, 9, 11}

- The value of the sum, m = 20
- Cross over rate: 3

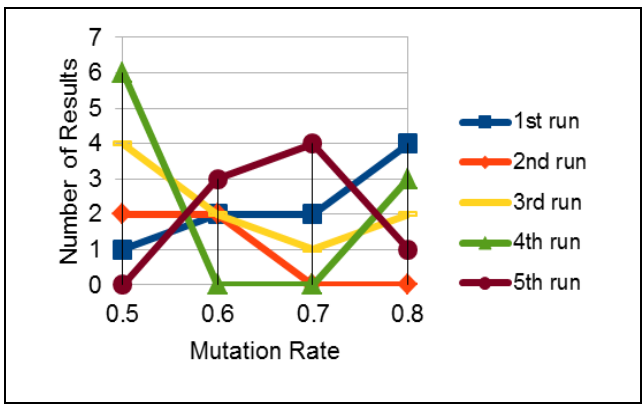

**Fig 7: Results of Experiment 6.**

Seventh Experiment

- Set A1 = {1, 3, 5, 7, 9, 11}
- The value of the sum, m = 20
- Cross over rate: 4

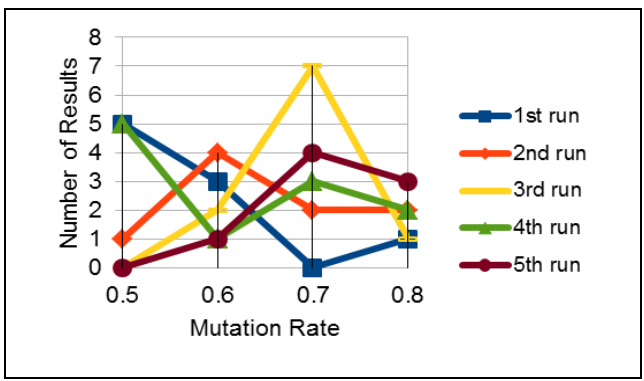

**Fig 8: Results of Experiment 7.**

Eighth Experiment

- Set A1 = {1, 3, 5, 7, 9, 11}
- The value of the sum, m = 20
- Cross over rate: 5

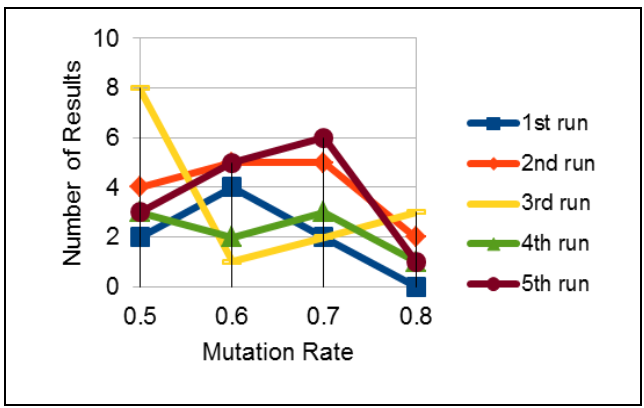

**Fig 9: Results of Experiment 8.**

Ninth Experiment

- Set A1 = {5, 7, 21, 33, 37, 91}
- The value of the sum, m = 112
- Cross over rate: 2

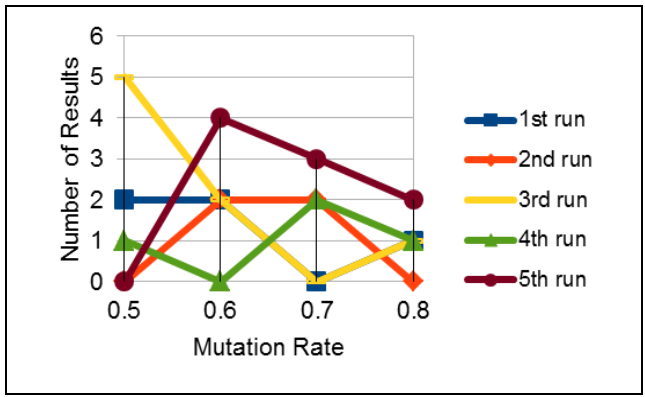

**Fig 10: Results of Experiment 9.**

Tenth Experiment

- Set A1 = {5, 7, 21, 33, 37, 91}
- The value of the sum, m = 112
- Cross over rate: 3

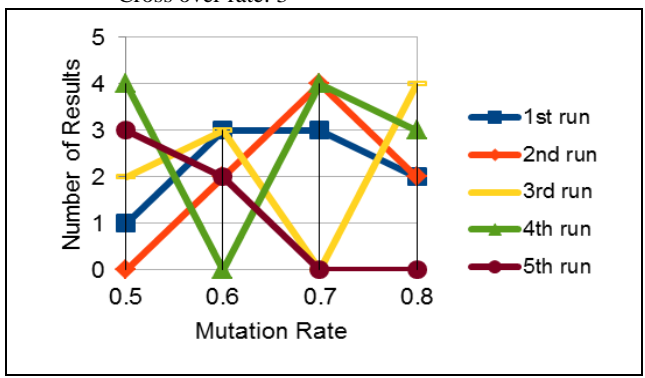

**Fig 11: Results of Experiment 10.**

Eleventh Experiment

- Set A1 = {5, 7, 21, 33, 37, 91}
- The value of the sum, m = 112
- Cross over rate: 4

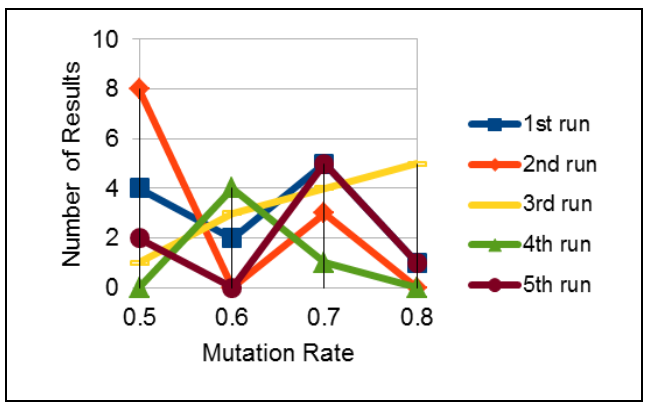

**Fig 12: Results of Experiment 11.**

Twelfth Experiment

- Set A1 = {5, 7, 21, 33, 37, 91}

- The value of the sum, m = 112
- Cross over rate: 5

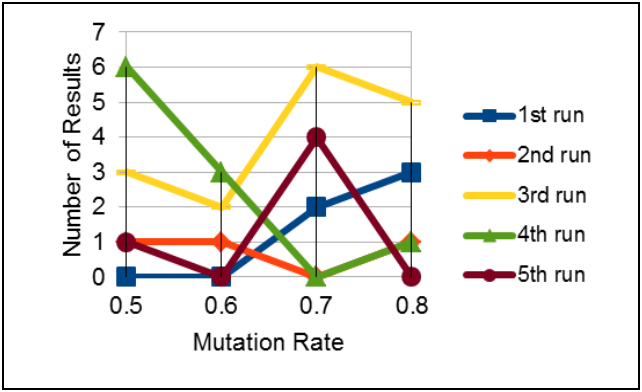

**Fig 13: Results of Experiment 12.**

Thirteenth Experiment

- Set A1 = {2, 9, 21, 33, 77, 101}
- The value of the sum, m = 79
- Cross over rate: 2

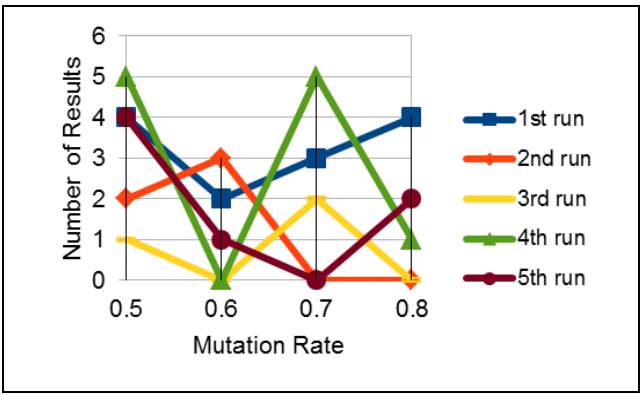

**Fig 14: Results of Experiment 13.**

Fourteenth Experiment

- Set A1 = {2, 9, 21, 33, 77, 101}
- The value of the sum, m = 79
- Cross over rate: 3

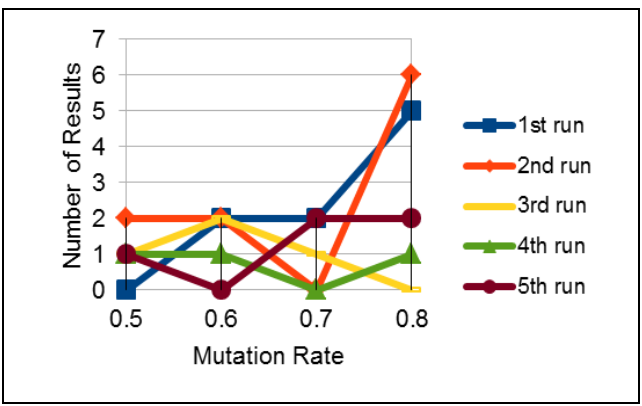

**Fig 15: Results of Experiment 14.**

Fifteenth Experiment

- Set A1 = {2, 9, 21, 33, 77, 101}
- The value of the sum, m = 79
- Cross over rate: 4

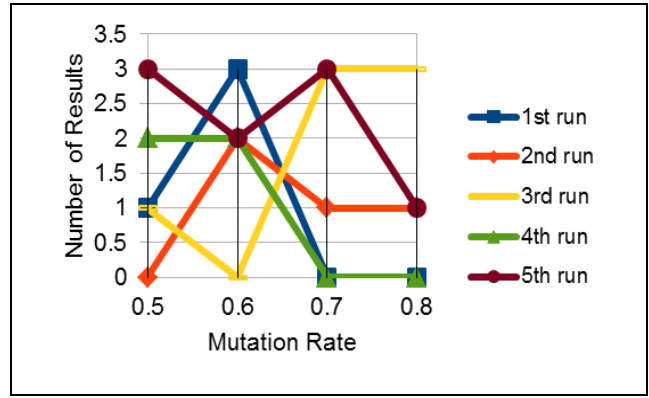

**Fig 16: Results of Experiment 15.**

Sixteenth Experiment

- Set A1 = {2, 9, 21, 33, 77, 101}
- The value of the sum, m = 79
- Cross over rate: 5

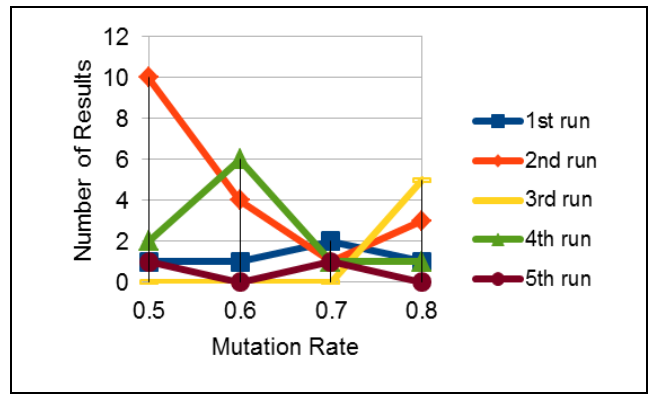

**Fig 17: Results of Experiment 16.**

Seventeenth Experiment

- Set A1 = {7, 10, 13, 20, 27, 30}
- The value of the sum, m = 57
- Cross over rate: 2

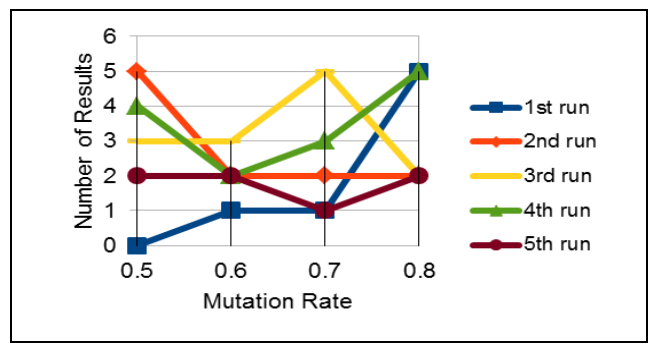

**Fig 18: Results of Experiment 17.**

Eighteenth Experiment

- Set A1 = {7, 10, 13, 20, 27, 30}
- The value of the sum, m = 57
- Cross over rate: 3

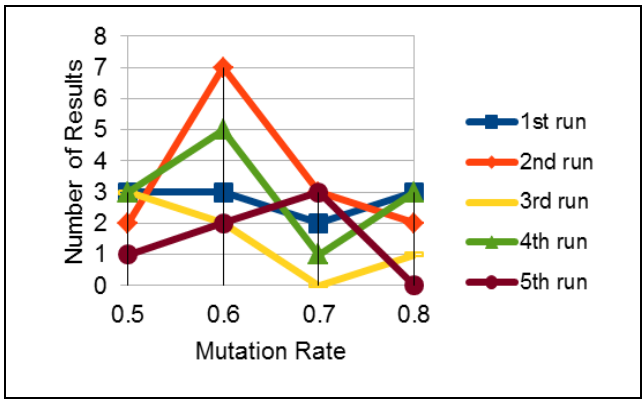

**Fig 19: Results of Experiment 18.**

Nineteenth Experiment

- Set A1 = {7, 10, 13, 20, 27, 30}
- The value of the sum, m = 57
- Cross over rate: 4

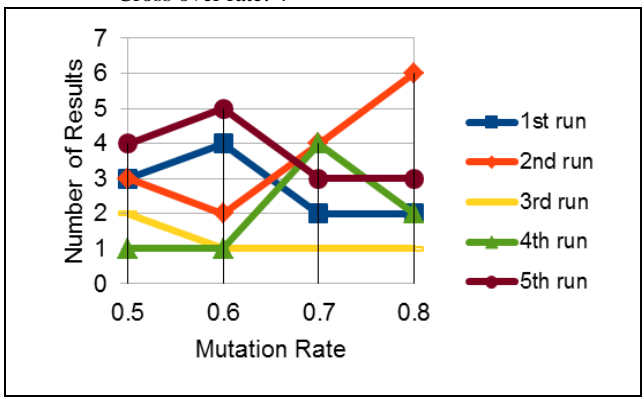

**Fig 20: Results of Experiment 19.**

Twentieth Experiment

- Set A1 = {7, 10, 13, 20, 27, 30}
- The value of the sum, m = 57
- Cross over rate: 5

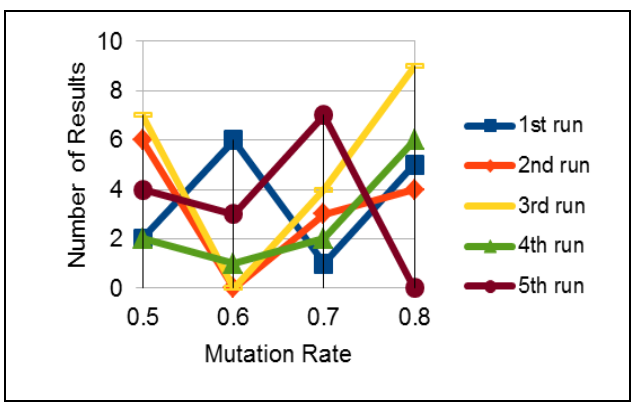

**Fig 21: Results of Experiment 20.**

Graph of number of solutions obtained with mutation rate is shown in above figures. The lines of different colors depict the results obtained in the successive runs of the algorithm. The results have been summarized in the next section.

## 6. CONCLUSIONS AND FUTURE SCOPE

The number of solutions obtained by caring out the experiments point to the fact that the method gives best results when the cross- over rate is 2 percent. It may also be noted that the theory of GAs also point to the same fact. If the crossover rate is too large then the probability of generation of chromosomes same as that of initial population would be high. The crossover rate, therefore, should neither be too high or too low. As per the results, if the mutation rate is 0.6 then the average of the number of solutions is maximized. So, the above method performs well if the crossover rate is 2 percent and the mutation rate is 0.6. It may also be stated here that, the total number of solutions is taken as the criteria of goodness. The above experiment is now being carried out using Diploid Genetic Algorithms in order to test the method in dynamic environments[22]. Moreover, the number of samples in the new experiment is increased to 100. The crossover rate ranges from 2-8 (with a step of 0.1) and the mutation rate from 0.01 to 1.0 (step = 0.01). Thus, the experiment would be carried out 7000 times. The present paper, however, proposes a new methodology, implements it and analyzes the results. The results would pave the way of application of variants of GAs and Artificial Life to cryptanalysis.

## 7. REFERENCES

[1] Schneier, B., "Applied Cryptography", in John Wiley & Sons, 1996.Ding, W. and Marchionini, G. 1997 A Study on Video Browsing Strategies. Technical Report. University of Maryland at College Park.

[2] Shamir, A., "How To Share A Secret", in Communications of the ACM, Volume 22 Issue 11, Nov. 1979.Tavel, P. 2007 Modeling and Simulation Design. AK Peters Ltd.

[3] Kitchenham, B., Mendes, E., et. al., (2007) "A Systematic Review of Cross- vs. Within-Company Cost Estimation Studies", IEEE Trans on SE, 33(5), pp 316-329.

[4] Engström, E., "A Systematic Review on Regression Test Selection Techniques", in Journal of Information and Software Technology 52(1):14-30, 2010.

[5] Sinha, S.N., "A cryptanalytic attack on Knapsack cipher using Differential Evolution algorithm", in Recent Advances in Intelligent Computational Systems (RAICS), 2011 IEEE.

[6] Raghuvamshi, A. And Rao, P.V., "An Effortless Cryptanalytic Attack on Knapsack Cipher" in Process Automation, Control and Computing (PACC), 2011 International Conference.

[7] Palit, S. et. al., "A cryptanalytic attack on the knapsack cryptosystem using binary Firefly algorithm", Computer and Communication Technology (ICCCT), 2011 2nd International Conference.

[8] AbdulHalim, M.F. and Sakhir, "A binary Particle Swarm Optimization for attacking knapsacks Cipher Algorithm", Computer and Communication Engineering, 2008. ICCCE 2008. International Conference.

[9] Nalini, N. and Raghavendra Rao, G., "Attacks of simple block ciphers via efficient heuristics", Information Sciences: an International Journal archive Volume 177 Issue 12, June, 2007, Elsevier Science Inc. New York, NY, USA.

[10] Nalini, N. and Raghavendra Rao, G., "Experiments on cryptanalysing block ciphers via evolutionary computation paradigms", EC'06 Proceedings of the 7th WSEAS International Conference on Evolutionary Computing, World Scientific and Engineering Academy and Society (WSEAS), Stevens Point, Wisconsin, USA ©2006.

[11] Song, J. et. al., "Cryptanalysis of two-round DES using genetic algorithms", GECCO '08 Proceedings of the 10th annual conference on Genetic and evolutionary computation, ACM New York, NY, USA ©2008.

[12] Song, J. et. al., "Cryptanalysis of Two-Round DES Using Genetic Algorithm", Advances in Computation and Intelligence Lecture Notes in Computer Science Volume 4683, 2007, pp 583-590.

[13] Preneel B., "Cryptographic Primitives for Information Authentication — State of the Art", State of the Art in Applied Cryptography Lecture Notes in Computer Science Volume 1528, 1998, pp 49-104.

[14] Millan W. et. al., "Heuristic design of cryptographically strong balanced Boolean functions", Advances in Cryptology — EUROCRYPT' 98 Lecture Notes in Computer Science Volume 1403, 1998, pp 489-499.

[15] Preneel B., "The State of Cryptographic Hash Functions", Lectures on Data Security Lecture Notes in Computer Science Volume 1561, 1999, pp 158-182.

[16] Preneel B., "The State of Hash Functions and the NIST SHA-3 Competition", Information Security and Cryptology Lecture Notes in Computer Science Volume 5487, 2009, pp 1-11.

[17] Burnett L. et. al., "Evolutionary Heuristics for Finding Cryptographically Strong S-Boxes", Information and Communication Security Lecture Notes in Computer Science Volume 1726, 1999, pp 263-274.

[18] Preneel B., "The First 30 Years of Cryptographic Hash Functions and the NIST SHA-3 Competition", Topics in Cryptology - CT-RSA 2010 Lecture Notes in Computer Science Volume 5985, 2010, pp 1-14.

[19] Kobayashi K. and Kimura M., "Improving the security of the knapsack cryptosystem", Electronics and Communications in Japan (Part III: Fundamental Electronic Science), Volume 80, Issue 7, pages 37–43, July 1997.

[20] Haitham Rashwan1 et. al., "Security of the GPT cryptosystem and its applications to cryptography", Security and Communication Networks, Volume 4, Issue 8, pages 937–946, August 2011.

[21] Lu1 R. et. al., "An efficient and provably secure public key encryption scheme based on coding theory", Security and Communication Networks, Volume 4, Issue 12, pages 1440–1447, December 2011.

[22] Bhasin H. et. al., On the applicability of diploid genetic algorithms in dynamic environments, pp 1-8, July 2015.

[23] Goldberg, Genetic Algorithms in Search, Optimization and Machine Learning.

[24] Bhasin H. et. al., "Cryptography Using genetic Algorithms", In Reliability Infocom Technology and Optimization Conference, 2010.

[25] Bhasin H. et. al., "Genetic Algorithm based solution to Maximum Clique Problem", International Journal of Computer Sceince and Engineering, Vol. 4, No. 8, 2012.

[26] Bhasin H. et. al., "Harnessing Cellular Automata and Genetic Algorithm to Solve Travelling Salesman Problem", In International Conference on Information, Computing and Telecommunications,(ICICT-2012), New Delhi, India.

[27] Bhasin H. et. al., "Modified Genetic Algorithm based Solution to Subset Sum problem", International Journal of Advanced Research in Artificial Intelligence, Vol. 1, No. 1, 2012.